\documentclass[twocolumn,aps]{revtex4}

\usepackage{graphicx}
\usepackage{dcolumn}
\usepackage{bm}
\usepackage{times}

\begin{document}
\title{On Magnetic Interlayer Coupling and Proximity Effect in a La$_{0.67}$Ca$_{0.33}$MnO$_3$(10 nm)/YBa$_2$Cu$_3$O$_7$(10 nm) Superlattice}

\author{J.W. Freeland$^1$, J. Chakhalian$^2$, H.-U. Habermeier$^{3}$, G. Cristiani$^3$ and B Keimer$^3$}
\affiliation{$^1$Advanced Photon Source, Argonne National Laboratory, Argonne, Illinois 60439, USA}
\affiliation{$^2$University of Arkansas, Fayetteville, Arkansas 72701, USA}
\affiliation{$^3$Max Planck Institute for Solid State Research, Stuttgart D-70561, Germany}

\begin{abstract} 
We present a study of interlayer coupling and proximity effects in a La$_{0.66}$Ca$_{0.33}$MnO$_3$(10 nm)/YBa$_2$Cu$_3$O$_7$(10 nm) superlattice. Using element-sensitive x-ray probes, the magnetic state of Mn can be probed  without seeing the strong diamagnetism of the superconductor, which makes this approach ideal to study changes in the magnetic properties across the superconducting transition. By a combined experiment using {\it in situ} transport measurements during polarized soft x-ray measurements, we were able to see no noticeable influence of the superconducting state on the magnetic properties and no evidence for magnetic coupling across a 10 nm YBCO layer. 
 \end{abstract} 
\maketitle
Artificial tailor-made superlattices (SL) composed of  alternating layers with mutually antagonistic order parameters have become an important test-ground for exploration of novel quantum states and unusual physical phenomena \cite{ahn:1185}.
For the case of ferromagnetic(FM)/superconducting(SC) junctions, there has been a tremendous amount of work on proximity effects to understand effects of coupling between the FM and SC order parameters (see Ref.\cite{buzdin:935,bergeret:1321} and references therein). It has been predicted that the presence of a SC layer next to the ferromagnet causes a suppression of the magnetic order, while the FM induces a small magnetic moment in the SC layer. This problem has two natural lengthscales, namely, the effect of magnetism in the SC is determined by the superconducting coherence length ($\xi_s$) whereas the behavior of superelectrons  in the in the FM is determined by the electron diffusion constant of the FM ($\xi_f$). Recently theoretical effort has been devoted to the interesting case of half-metallic FM/high-temperature SC interfaces\cite{lofwander:187003} with particular attention to the most representative materials for  each class, i.e.  La$_{0.7}$Ca$_{0.3}$MnO$_3$ (LCMO)  and YBa$_2$Cu$_3$O$_7$ (YBCO). 

Several groups have attempted experimental tests of the theory by studying  c-axis-grown superlattices composed of YBCO and LCMO layers.  In recent extensive studies, Sefrioui et al. and  Pena et al. concluded that leakage of magnetism into the YBCO layer was small due to the short SC coherence length along the c-dxis ($\le$1 nm) and attributed a strong suppression of the magnetic order in the LCMO layer to a long range proximity effect \cite{sefrioui:214511,pena:224502}. On the other hand, Senapati and Budhani reported a non-oscillatory anti-ferromagnetic coupling across the YBCO layer in both the normal and superconducting state \cite{senapati:224507}. These results are unexpected and still unexplained theoretically. A number of phenomenological models have been proposed to account for the long-range proximity effect including the formation of zero-energy bound states which may dominate the SC/FM transport\cite{stefanakis}, penetration of SC pairs into the half-metallic  FM layer through the domain walls  where the exchange field is considerably reduced \cite{melin}, and by virtue of the penetration of SC triplet correlations into the FM layer[1,2]. However, none of the models could completely  account for the observed proximity phenomena.  Hence, microscopic measurements on FM/SC SLs with enhanced sensitivity to magnetism  at the nanoscale are required.
\begin{figure}[h]
\begin{tabular}{c}
	\includegraphics[scale=.4]{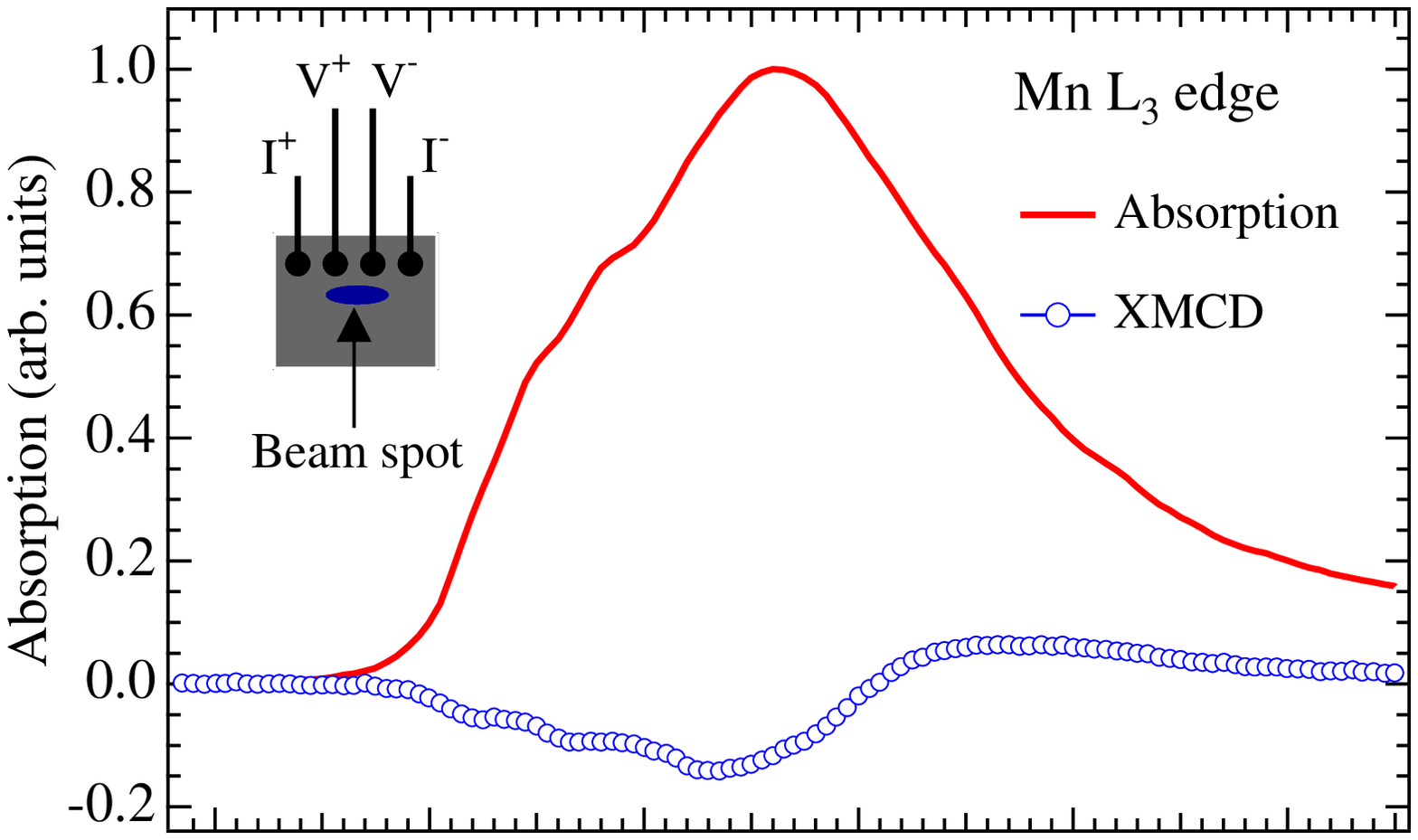} \\
	\includegraphics[scale=.4]{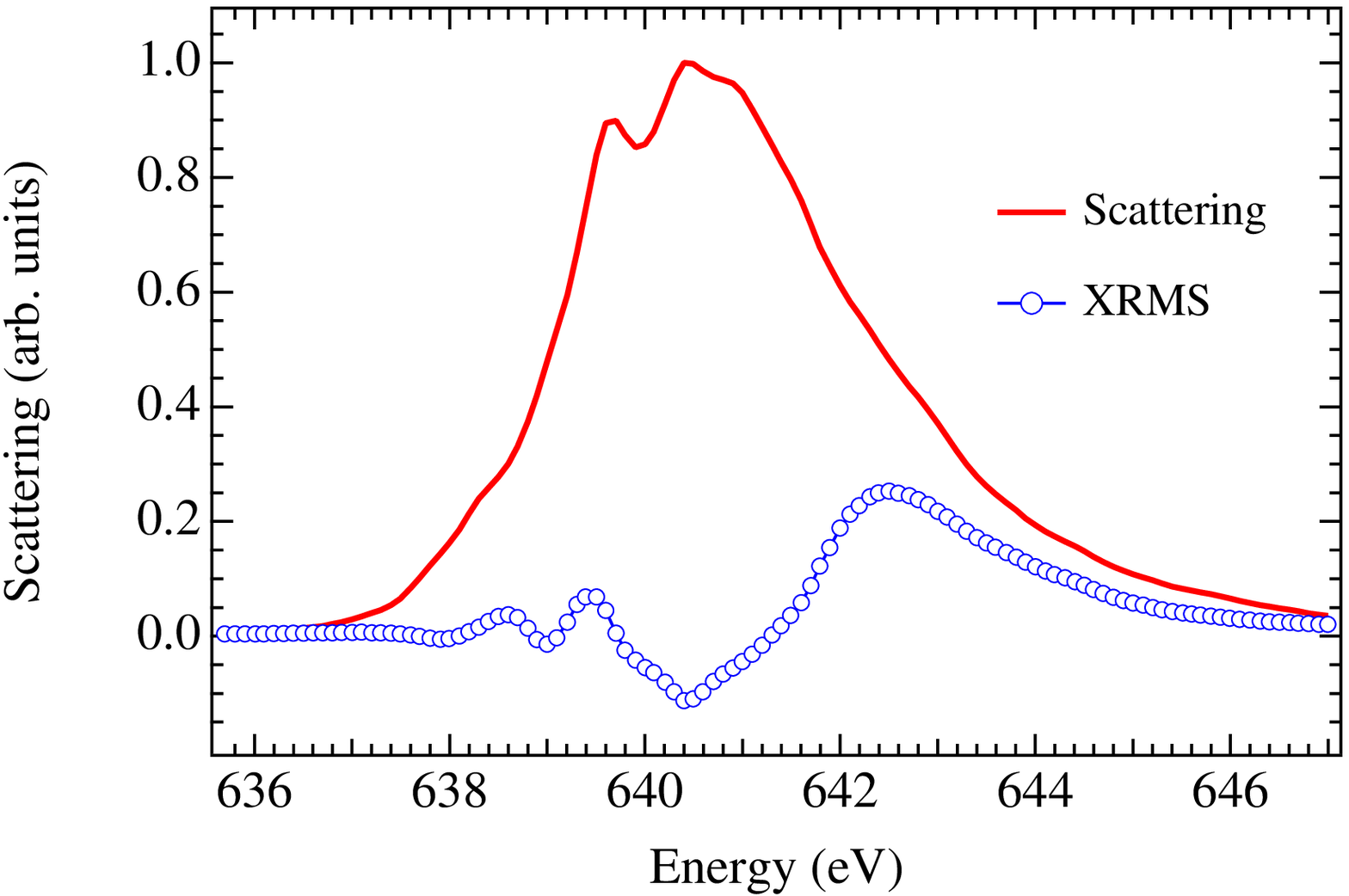} \\
\end{tabular}
\caption{(Top) Absorption and XMCD measured at the Mn L$_3$ edge at T = 65 K. Inset shows the 4-point transport geometry for simultaneous measurement of x-ray interactions and sample resistance. (Bottom) Scattering and XRMS measured near the (003) Bragg peak (incidence angle $\sim$ 11 deg.).}
\protect\label{xmcdxrms} 
\end{figure}

In this letter, we present results from chemical element-sensitive x-ray measurements of the magnetic properties of a  representative  La$_{0.66}$Ca$_{0.33}$MnO$_3$(10 nm)/YBa$_2$Cu$_3$O$_7$(10 nm) superlattice. Since the x-ray probe is not affected by the diamagnetism of  the superconductor, this technique is well suited to investigate the interlayer coupling and proximity effect across the superconducting transition. In a combined experiment,  {\it in situ} transport probes were obtained concomitant with x-ray magnetic circular dichroism (XMCD) and x-ray resonant magnetic scattering (XRMS) spectra taken at the L$_{3,2}$ Mn edges. We concluded that there was no noticeable influence of the superconducting order parameter on the magnetic properties  and no evidence for a magnetic coupling across a 10 nm thick YBCO layer.  

The microscopic magnetic properties of the superlattices were studied by polarized x-ray techniques at beamline 4-ID-C of the Advanced Photon Source \cite{blrsi}. The experimental configuration allows for simultaneous measurement of resonant x-ray absorption and scattering by switching the polarization between left-circular polarization (LCP) and right-circular polarization (RCP).   The sum (I$^+$+I$^-$) of these signals provides information on the electronic environment of the Mn 3$\it{d}$ electrons while magnetic information is contained in the difference (I$^+$-I$^-$), which are x-ray magnetic circular dichroism (XMCD) \cite{xmcd} and x-ray resonant magnetic scattering(XRMS) \cite{xrms,kortright:12216}, respectively (see Fig.\ \ref{xmcdxrms}).  The data were obtained simultaneously in two complimentary modes characterized by the very different depth sensitivity. Measurements of absorption by monitoring the photocurrent due to escaping electrons, referred to as  total electron yield (TEY), probe primarily the magnetic order in the top LCMO layer and in the vicinity of the first interface of the superlattice. Unlike TEY, XRMS is more bulk sensitive and is able to deliver information from several interfaces. Moreover, since the XRMS signal is directly proportional to the sample magnetization, scattering was used to obtain the magneto-optical  hysteresis loops. For these measurements, the sample angle was tuned to near the (003) superlattice Bragg peak (incidence angle $\sim$ 11 deg.), which, according our XRMS computer simulations, is sensitive to the magnetization averaged over several superlattice repeats.

\begin{figure}[h]
\begin{tabular}{c}
	\includegraphics[scale=.4]{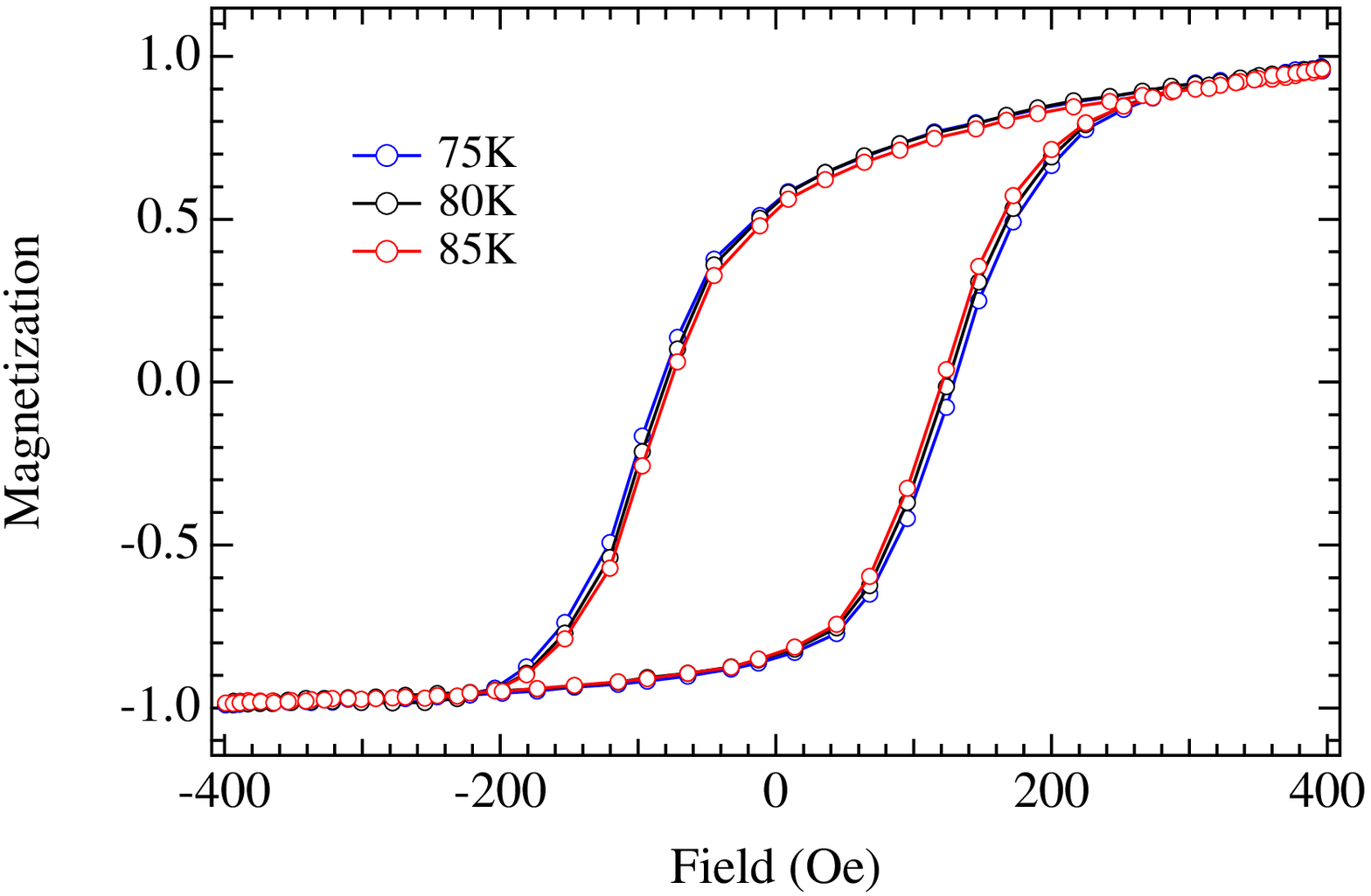} \\
	\includegraphics[scale=.4]{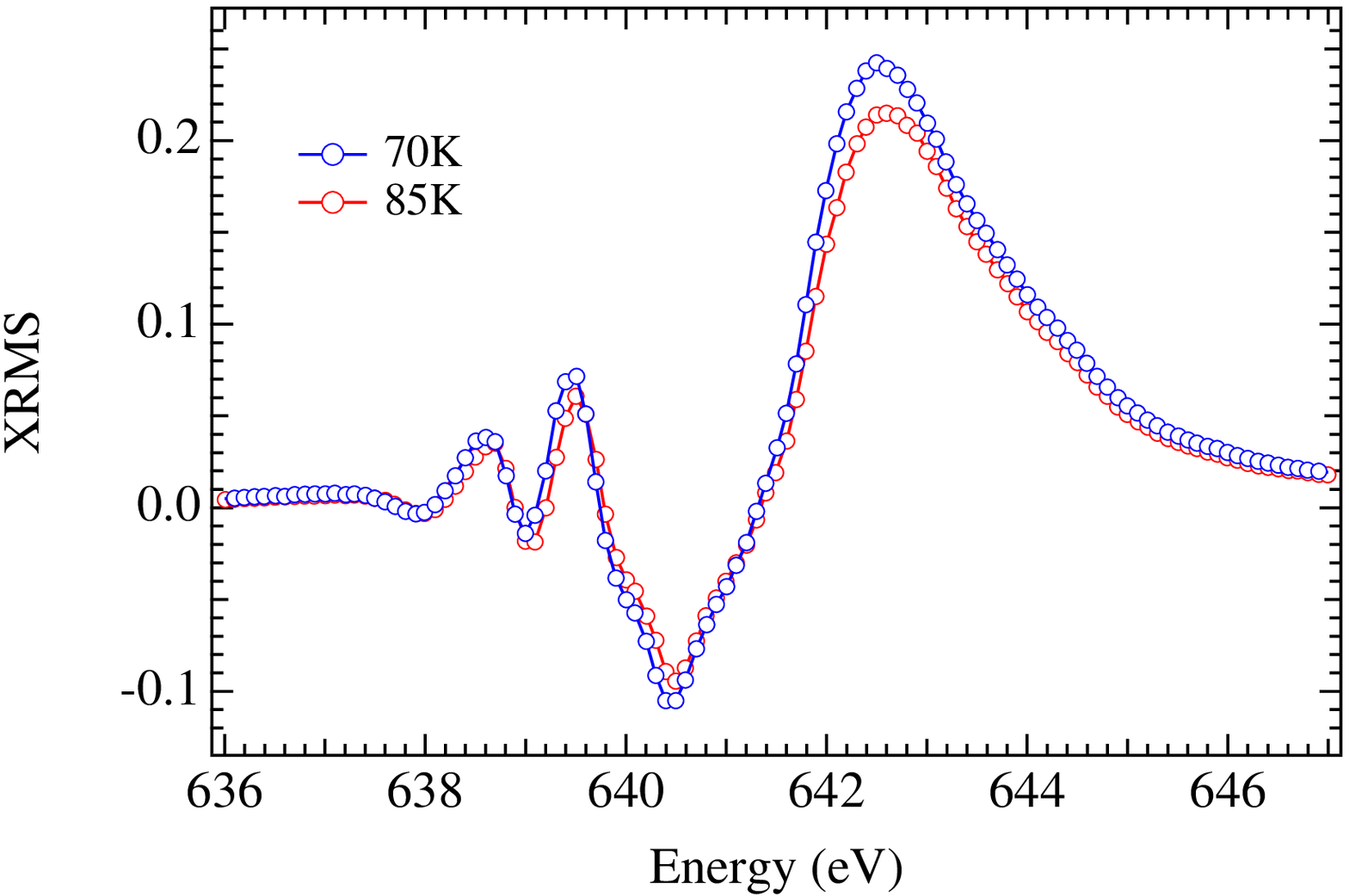} \\
\end{tabular}
\caption{(Top) Magnetic hysteresis measured across the superconducting transition. (Bottom) X-ray resonant magnetic scattering at the Mn L$_3$ edge above and below the superconducting transition.}
\protect\label{hys} 
\end{figure} 

To measure the sample resistance  we attached indium contacts arranged in a 4-point geometry (see Fig.\ \ref{xmcdxrms}). Using a current of 10 $\mu$A, we were able to collect {\it in situ} transport data simultaneously with x-ray data at each temperature point. Aside from a small current offset, the transport current showed no observable effect on the measurement of the TEY via the sample drain current. 
Samples were grown using pulsed laser deposition (PLD) on 5x5x0.5 mm$^3$ SrTiO$_3$ (001) substrates \cite{growth,holden:064505,stahn:140509}. Their high quality was confirmed by x-ray diffraction, which showed epitaxial growth with the c-axis along (001). Resistivity and superconducting quantum interference device (SQUID) magnetization measurements revealed a FM transition at T$_{\rm{mag}}\sim$180K and a superconducting transition at T$_{\rm{sc}}$ = 75K. This is close to the transition temperature of the target material indicating good valency transfer. 

\begin{figure}[h]
\begin{tabular}{c}
	\includegraphics[scale=.4]{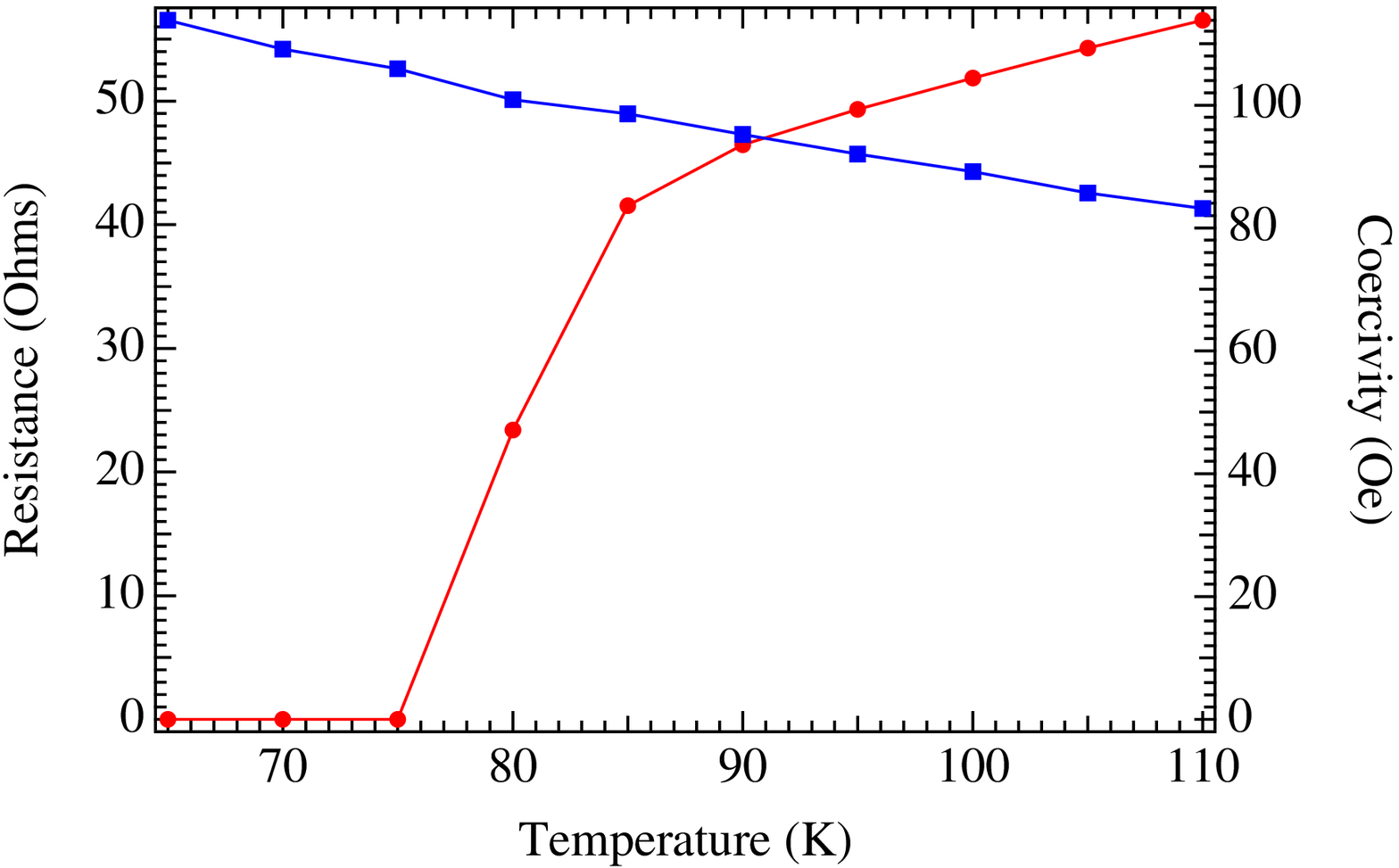}\\
	\includegraphics[scale=.4]{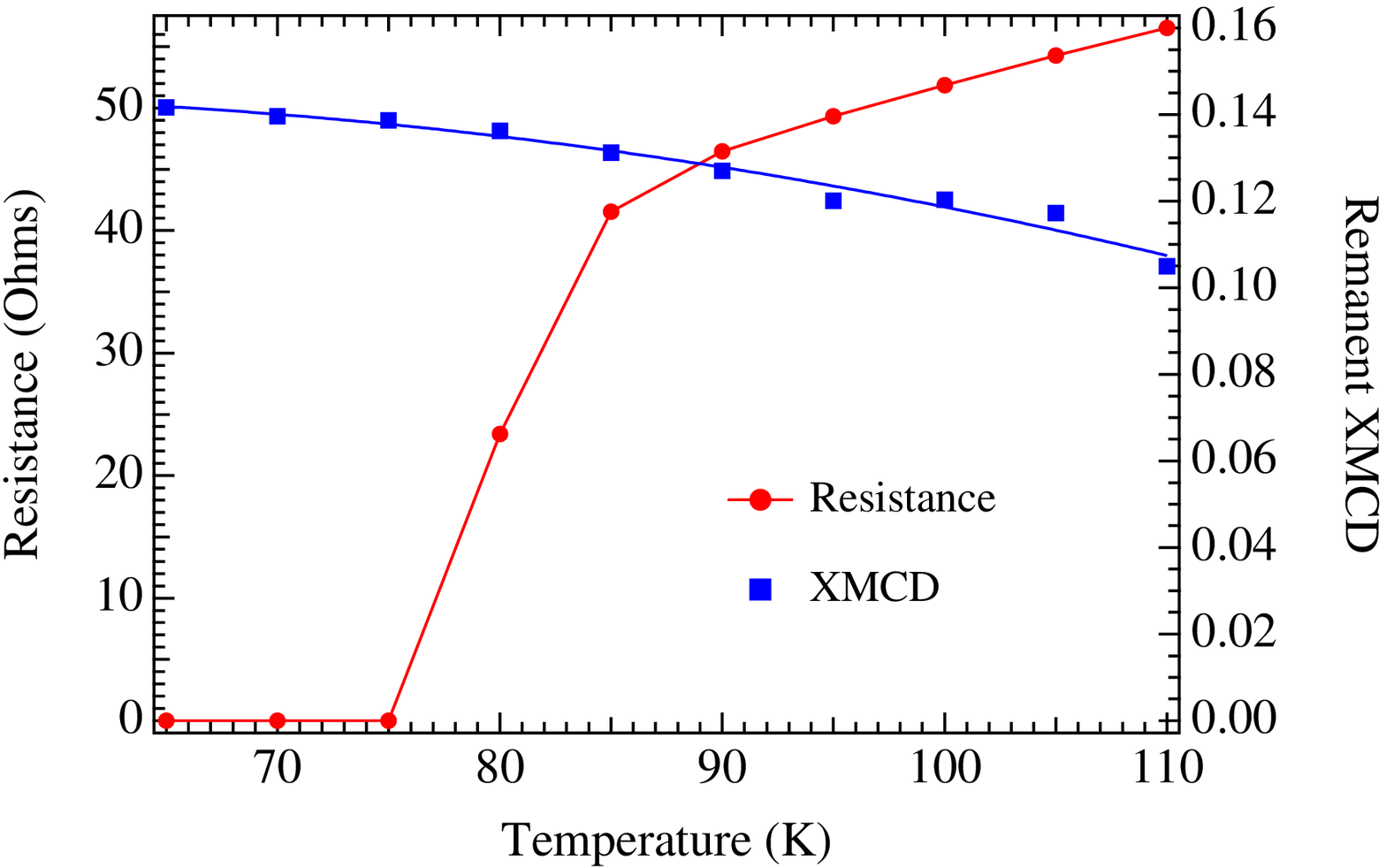} \\
\end{tabular}
\caption{(Top) Temperature dependence of sample resistance and magnetic coercivity across the superconducting transition. (Bottom) Temperature dependence of sample resistance and remanent XMCD signal across the superconducting transition. }
\protect\label{xmcdrvst} 
\end{figure}

First we discuss  the effect of interlayer coupling. It is well known that in the case of metallic superlattices, the most pronounced manifestation of the interlayer coupling is its effect on the magnetization reversal. If the film possesses strong antiferromagnetic (AFM) coupling, then pronounced plateaus will appear in the hysteresis loop when the magnetic moments of the layers are antiparallel, which, for an even number of equivalent layers, would result in a zero magnetization state. The top panel of Figure \ \ref{hys} shows the hysteresis loops taken at the Mn edge across the superconducting transition. As seen, the loops are  consistent with uniform layer reversal. 
There are only minor differences with increasing temperature which are due to the monotonic temperature dependence of hysteresis expected in any ferromagnetic material. The slight asymmetry of the loops results from interference of the scattered x-rays and is not related to an asymmetry in the magnetic reversal.  Together with the extracted magnetic coercivity in the top panel of Fig.\ \ref{xmcdrvst}, the absence of discontinuous behavior at T$_{\rm{sc}}$ clearly indicates the lack of an abrupt change which otherwise  would be present if there were significant coupling between the two LCMO layers across the YBCO spacer.  

Now we turn our attention to the long-range effects  of superelectrons on the FM layer. 
To search for evidence of the superconducting proximity effect inside the LCMO layer, we examined the behavior of the remanent magnetization in the top LCMO layer after field cooling in an applied field of H = +500 Oe (see Fig.\ \ref{xmcdrvst}). Theoretically it has been suggested that below $T_{\rm{sc}}$ leakage of superconductivity into the LCMO layer results in a suppression of the magnetization in the interface region, which is recovered with the loss of coherence upon heating above T$_{\rm{sc}}$. The data in the bottom panel of Fig.\ \ref{xmcdrvst} show no discontinuity at the superconducting transition temperature and are hence inconsistent with this scenario.In applied magnetic fields, we measured XMCD signals of $\sim35\%$ even in  the superconducting state, which is close to the value measured on a 10-nm LCMO thin film, indicating very little suppression of the magnetism in LCMO in the FM/SC superlattice. 

 To further verify this,  we probed the magnetization profile by measuring  the energy-dependent XRMS near the (003) Bragg reflection (see bottom panel of Fig.\ \ref{hys}). Our previous work has shown that the line shape of the energy dependence can be directly connected to the magnetic profile across the superlattice \cite{natmat,jpcm}. As evident from the bottom panel in Fig.\ \ref{hys} the only change in the lineshape across the superconducting transition is a slight decrease in overall amplitude due to a drop in the sample magnetization as the transition temperature is approached.  From this we conclude that  there is no change in the magnetization profile across the LCMO layer. Recent measurements of a magnetization profile in YBCO/LCMO  by neutron reflectivity do show a reduced LCMO magnetization in the interface region, but this persists even above T$_{\rm{sc}}$ \cite{stahn:140509}. This is not necessarily a consequence of the presence of YBCO, but may rather be due to the fragile nature of magnetic order at manganite interfaces \cite{jpcm}.

To explain these results we refer our previous microscopic investigation  of the magnetism at the interface between YBCO and LCMO.  
A combination of diffuse neutron reflectivity and x-ray absorption revealed  the reduced magnetic moment on Mn and the presence of  a net  magnetic moment on Cu, which was largely attributed to the substantially modified atomic and electronic structure at the interface \cite{natphys,stahn:140509}.  More specifically, because Cu chains at the YBCO/LCMO interface are missing \cite{varela} and because the chains are the charge reservoir essential for superconductivity,  the truncated first unit cell is likely non-superconducting and in the AFM state.  Coupling to the ferromagnetic LCMO at the interface cants the spins in this layer to result in the non-zero XMCD seen for Cu \cite{natphys}. This in turn serves to reduce the direct coupling of the superconducting regions in the YBCO layer with the LCMO layer. Due to the short c-axis coherence length, it would be hard to transfer the superconducting state into the LCMO across the magnetic region in the YBCO. Even if this were possible, a short $\xi_f \sim 0.2 nm$ is expected from the measured Fermi velocity of LCMO \cite{sefrioui:214511}.
 
 To summarize, we have reported the chemically sensitive soft x-ray XMCD measurements of the magnetic properties of YBCO(10 nm)/LCMO(10 nm) superlattices correlated with {\it in situ} transport measurements. The magnetic data taken at the L$_{3,2}$ Mn edge show the absence of any substantial changes around the superconducting transition and thus exclude a conventional microscopic coupling between the YBCO layers through the LCMO spacer. Measurements of the magnetization reversal show only coherent rotation of the layers and no sign of strong magnetic coupling across the 10 nm thick YBCO layer.

Two of the authors (J.W.F. and J.C.) would like to acknowledge very fruitful discussions with Matthias Eschrig. Work at Argonne is supported by the U.S. Department of Energy, Office of Science, under Contract No.\  DE-AC02-06CH11357.


\begin{thebibliography}{20}
\expandafter\ifx\csname natexlab\endcsname\relax\def\natexlab#1{#1}\fi
\expandafter\ifx\csname bibnamefont\endcsname\relax
  \def\bibnamefont#1{#1}\fi
\expandafter\ifx\csname bibfnamefont\endcsname\relax
  \def\bibfnamefont#1{#1}\fi
\expandafter\ifx\csname citenamefont\endcsname\relax
  \def\citenamefont#1{#1}\fi
\expandafter\ifx\csname url\endcsname\relax
  \def\url#1{\texttt{#1}}\fi
\expandafter\ifx\csname urlprefix\endcsname\relax\def\urlprefix{URL }\fi
\providecommand{\bibinfo}[2]{#2}
\providecommand{\eprint}[2][]{\url{#2}}

\bibitem[{\citenamefont{Ahn et~al.}(2006)\citenamefont{Ahn, Bhattacharya,
  Ventra, Eckstein, Frisbie, Gershenson, Goldman, Inoue, Mannhart, Millis
  et~al.}}]{ahn:1185}
\bibinfo{author}{\bibfnamefont{C.~H.} \bibnamefont{Ahn}},
  \bibinfo{author}{\bibfnamefont{A.}~\bibnamefont{Bhattacharya}},
  \bibinfo{author}{\bibfnamefont{M.~D.} \bibnamefont{Ventra}},
  \bibinfo{author}{\bibfnamefont{J.~N.} \bibnamefont{Eckstein}},
  \bibinfo{author}{\bibfnamefont{C.~D.} \bibnamefont{Frisbie}},
  \bibinfo{author}{\bibfnamefont{M.~E.} \bibnamefont{Gershenson}},
  \bibinfo{author}{\bibfnamefont{A.~M.} \bibnamefont{Goldman}},
  \bibinfo{author}{\bibfnamefont{I.~H.} \bibnamefont{Inoue}},
  \bibinfo{author}{\bibfnamefont{J.}~\bibnamefont{Mannhart}},
  \bibinfo{author}{\bibfnamefont{A.~J.} \bibnamefont{Millis}},
  \bibnamefont{et~al.}, \bibinfo{journal}{Reviews of Modern Physics}
  \textbf{\bibinfo{volume}{78}}, \bibinfo{eid}{1185} (\bibinfo{year}{2006}).

\bibitem[{\citenamefont{Buzdin}(2005)}]{buzdin:935}
\bibinfo{author}{\bibfnamefont{A.~I.} \bibnamefont{Buzdin}},
  \bibinfo{journal}{Reviews of Modern Physics} \textbf{\bibinfo{volume}{77}},
  \bibinfo{eid}{935} (\bibinfo{year}{2005}).

\bibitem[{\citenamefont{Bergeret et~al.}(2005)\citenamefont{Bergeret, Volkov,
  and Efetov}}]{bergeret:1321}
\bibinfo{author}{\bibfnamefont{F.~S.} \bibnamefont{Bergeret}},
  \bibinfo{author}{\bibfnamefont{A.~F.} \bibnamefont{Volkov}},
  \bibnamefont{and} \bibinfo{author}{\bibfnamefont{K.~B.}
  \bibnamefont{Efetov}}, \bibinfo{journal}{Reviews of Modern Physics}
  \textbf{\bibinfo{volume}{77}}, \bibinfo{eid}{1321} (\bibinfo{year}{2005}).

\bibitem[{\citenamefont{Lofwander et~al.}(2005)\citenamefont{Lofwander,
  Champel, Durst, and Eschrig}}]{lofwander:187003}
\bibinfo{author}{\bibfnamefont{T.}~\bibnamefont{Lofwander}},
  \bibinfo{author}{\bibfnamefont{T.}~\bibnamefont{Champel}},
  \bibinfo{author}{\bibfnamefont{J.}~\bibnamefont{Durst}}, \bibnamefont{and}
  \bibinfo{author}{\bibfnamefont{M.}~\bibnamefont{Eschrig}},
  \bibinfo{journal}{Physical Review Letters} \textbf{\bibinfo{volume}{95}},
  \bibinfo{eid}{187003} (\bibinfo{year}{2005}).

\bibitem[{\citenamefont{Sefrioui et~al.}(2003)\citenamefont{Sefrioui, Arias,
  Pena, Villegas, Varela, Prieto, Leon, Martinez, and
  Santamaria}}]{sefrioui:214511}
\bibinfo{author}{\bibfnamefont{Z.}~\bibnamefont{Sefrioui}},
  \bibinfo{author}{\bibfnamefont{D.}~\bibnamefont{Arias}},
  \bibinfo{author}{\bibfnamefont{V.}~\bibnamefont{Pena}},
  \bibinfo{author}{\bibfnamefont{J.~E.} \bibnamefont{Villegas}},
  \bibinfo{author}{\bibfnamefont{M.}~\bibnamefont{Varela}},
  \bibinfo{author}{\bibfnamefont{P.}~\bibnamefont{Prieto}},
  \bibinfo{author}{\bibfnamefont{C.}~\bibnamefont{Leon}},
  \bibinfo{author}{\bibfnamefont{J.~L.} \bibnamefont{Martinez}},
  \bibnamefont{and}
  \bibinfo{author}{\bibfnamefont{J.}~\bibnamefont{Santamaria}},
  \bibinfo{journal}{Physical Review B} \textbf{\bibinfo{volume}{67}},
  \bibinfo{eid}{214511} (\bibinfo{year}{2003}).

\bibitem[{\citenamefont{Pena et~al.}(2004)\citenamefont{Pena, Sefrioui, Arias,
  Leon, Santamaria, Varela, Pennycook, and Martinez}}]{pena:224502}
\bibinfo{author}{\bibfnamefont{V.}~\bibnamefont{Pena}},
  \bibinfo{author}{\bibfnamefont{Z.}~\bibnamefont{Sefrioui}},
  \bibinfo{author}{\bibfnamefont{D.}~\bibnamefont{Arias}},
  \bibinfo{author}{\bibfnamefont{C.}~\bibnamefont{Leon}},
  \bibinfo{author}{\bibfnamefont{J.}~\bibnamefont{Santamaria}},
  \bibinfo{author}{\bibfnamefont{M.}~\bibnamefont{Varela}},
  \bibinfo{author}{\bibfnamefont{S.~J.} \bibnamefont{Pennycook}},
  \bibnamefont{and} \bibinfo{author}{\bibfnamefont{J.~L.}
  \bibnamefont{Martinez}}, \bibinfo{journal}{Physical Review B}
  \textbf{\bibinfo{volume}{69}}, \bibinfo{eid}{224502} (\bibinfo{year}{2004}).

\bibitem[{\citenamefont{Senapati and Budhani}(2005)}]{senapati:224507}
\bibinfo{author}{\bibfnamefont{K.}~\bibnamefont{Senapati}} \bibnamefont{and}
  \bibinfo{author}{\bibfnamefont{R.~C.} \bibnamefont{Budhani}},
  \bibinfo{journal}{Physical Review B} \textbf{\bibinfo{volume}{71}},
  \bibinfo{eid}{224507} (\bibinfo{year}{2005}).

\bibitem[{\citenamefont{Stefanakis and M\'{e}lin}(2003)}]{stefanakis}
\bibinfo{author}{\bibfnamefont{N.}~\bibnamefont{Stefanakis}} \bibnamefont{and}
  \bibinfo{author}{\bibfnamefont{R.}~\bibnamefont{M\'{e}lin}},
  \bibinfo{journal}{Journal of Physics: Condensed Matter}
  \textbf{\bibinfo{volume}{15}}, \bibinfo{pages}{4239} (\bibinfo{year}{2003}).

\bibitem[{\citenamefont{M\'elin and Peysson}(2003)}]{melin}
\bibinfo{author}{\bibfnamefont{R.}~\bibnamefont{M\'elin}} \bibnamefont{and}
  \bibinfo{author}{\bibfnamefont{S.}~\bibnamefont{Peysson}},
  \bibinfo{journal}{Phys. Rev. B} \textbf{\bibinfo{volume}{68}},
  \bibinfo{pages}{174515} (\bibinfo{year}{2003}).

\bibitem[{\citenamefont{Freeland et~al.}(2001)\citenamefont{Freeland, Lang,
  Srajer, Winarski, Shu, and Mills}}]{blrsi}
\bibinfo{author}{\bibfnamefont{J.~W.} \bibnamefont{Freeland}},
  \bibinfo{author}{\bibfnamefont{J.~C.} \bibnamefont{Lang}},
  \bibinfo{author}{\bibfnamefont{G.}~\bibnamefont{Srajer}},
  \bibinfo{author}{\bibfnamefont{R.}~\bibnamefont{Winarski}},
  \bibinfo{author}{\bibfnamefont{D.}~\bibnamefont{Shu}}, \bibnamefont{and}
  \bibinfo{author}{\bibfnamefont{D.~M.} \bibnamefont{Mills}},
  \bibinfo{journal}{Rev.\ Sci.\ Instrum.} \textbf{\bibinfo{volume}{73}},
  \bibinfo{pages}{1408} (\bibinfo{year}{2001}).

\bibitem[{\citenamefont{Chen et~al.}(1995)\citenamefont{Chen, Idzerda, {H.-J
  Lin}, {N.V. Smith}, Meigs, Chaban, {G.H. Ho}, Pellegrin, and Sette}}]{xmcd}
\bibinfo{author}{\bibfnamefont{C.~T.} \bibnamefont{Chen}},
  \bibinfo{author}{\bibfnamefont{Y.~U.} \bibnamefont{Idzerda}},
  \bibinfo{author}{\bibnamefont{{H.-J Lin}}},
  \bibinfo{author}{\bibnamefont{{N.V. Smith}}},
  \bibinfo{author}{\bibfnamefont{G.}~\bibnamefont{Meigs}},
  \bibinfo{author}{\bibfnamefont{E.}~\bibnamefont{Chaban}},
  \bibinfo{author}{\bibnamefont{{G.H. Ho}}},
  \bibinfo{author}{\bibfnamefont{E.}~\bibnamefont{Pellegrin}},
  \bibnamefont{and} \bibinfo{author}{\bibfnamefont{F.}~\bibnamefont{Sette}},
  \bibinfo{journal}{Phys.\ Rev.\ Lett.} \textbf{\bibinfo{volume}{75}},
  \bibinfo{pages}{152} (\bibinfo{year}{1995}).

\bibitem[{\citenamefont{Kao et~al.}(1990)\citenamefont{Kao, Hastings, Johnson,
  Siddons, and Smith}}]{xrms}
\bibinfo{author}{\bibfnamefont{C.~C.} \bibnamefont{Kao}},
  \bibinfo{author}{\bibfnamefont{J.~B.} \bibnamefont{Hastings}},
  \bibinfo{author}{\bibfnamefont{E.~D.} \bibnamefont{Johnson}},
  \bibinfo{author}{\bibfnamefont{D.~P.} \bibnamefont{Siddons}},
  \bibnamefont{and} \bibinfo{author}{\bibfnamefont{G.~C.} \bibnamefont{Smith}},
  \bibinfo{journal}{Phys.\ Rev.\ Lett.} \textbf{\bibinfo{volume}{65}},
  \bibinfo{pages}{373} (\bibinfo{year}{1990}).

\bibitem[{\citenamefont{Kortright and Kim}(2000)}]{kortright:12216}
\bibinfo{author}{\bibfnamefont{J.~B.} \bibnamefont{Kortright}}
  \bibnamefont{and} \bibinfo{author}{\bibfnamefont{S.-K.} \bibnamefont{Kim}},
  \bibinfo{journal}{Physical Review B} \textbf{\bibinfo{volume}{62}},
  \bibinfo{pages}{12216} (\bibinfo{year}{2000}).

\bibitem[{\citenamefont{Habermeier et~al.}(2001)\citenamefont{Habermeier,
  Cristiani, Kremer, Lebedev, and Tendeloo}}]{growth}
\bibinfo{author}{\bibfnamefont{H.-U.} \bibnamefont{Habermeier}},
  \bibinfo{author}{\bibfnamefont{G.}~\bibnamefont{Cristiani}},
  \bibinfo{author}{\bibfnamefont{R.~K.} \bibnamefont{Kremer}},
  \bibinfo{author}{\bibfnamefont{O.}~\bibnamefont{Lebedev}}, \bibnamefont{and}
  \bibinfo{author}{\bibfnamefont{G.~V.} \bibnamefont{Tendeloo}},
  \bibinfo{journal}{Physica C} \textbf{\bibinfo{volume}{354}},
  \bibinfo{pages}{29} (\bibinfo{year}{2001}).

\bibitem[{\citenamefont{Holden et~al.}(2004)\citenamefont{Holden, Habermeier,
  Cristiani, Golnik, Boris, Pimenov, Humlicek, Lebedev, Tendeloo, Keimer
  et~al.}}]{holden:064505}
\bibinfo{author}{\bibfnamefont{T.}~\bibnamefont{Holden}},
  \bibinfo{author}{\bibfnamefont{H.-U.} \bibnamefont{Habermeier}},
  \bibinfo{author}{\bibfnamefont{G.}~\bibnamefont{Cristiani}},
  \bibinfo{author}{\bibfnamefont{A.}~\bibnamefont{Golnik}},
  \bibinfo{author}{\bibfnamefont{A.}~\bibnamefont{Boris}},
  \bibinfo{author}{\bibfnamefont{A.}~\bibnamefont{Pimenov}},
  \bibinfo{author}{\bibfnamefont{J.}~\bibnamefont{Humlicek}},
  \bibinfo{author}{\bibfnamefont{O.~I.} \bibnamefont{Lebedev}},
  \bibinfo{author}{\bibfnamefont{G.~V.} \bibnamefont{Tendeloo}},
  \bibinfo{author}{\bibfnamefont{B.}~\bibnamefont{Keimer}},
  \bibnamefont{et~al.}, \bibinfo{journal}{Physical Review B}
  \textbf{\bibinfo{volume}{69}}, \bibinfo{eid}{064505} (\bibinfo{year}{2004}).

\bibitem[{\citenamefont{Stahn et~al.}(2005)\citenamefont{Stahn, Chakhalian,
  Niedermayer, Hoppler, Gutberlet, Voigt, Treubel, Habermeier, Cristiani,
  Keimer et~al.}}]{stahn:140509}
\bibinfo{author}{\bibfnamefont{J.}~\bibnamefont{Stahn}},
  \bibinfo{author}{\bibfnamefont{J.}~\bibnamefont{Chakhalian}},
  \bibinfo{author}{\bibfnamefont{C.}~\bibnamefont{Niedermayer}},
  \bibinfo{author}{\bibfnamefont{J.}~\bibnamefont{Hoppler}},
  \bibinfo{author}{\bibfnamefont{T.}~\bibnamefont{Gutberlet}},
  \bibinfo{author}{\bibfnamefont{J.}~\bibnamefont{Voigt}},
  \bibinfo{author}{\bibfnamefont{F.}~\bibnamefont{Treubel}},
  \bibinfo{author}{\bibfnamefont{H.-U.} \bibnamefont{Habermeier}},
  \bibinfo{author}{\bibfnamefont{G.}~\bibnamefont{Cristiani}},
  \bibinfo{author}{\bibfnamefont{B.}~\bibnamefont{Keimer}},
  \bibnamefont{et~al.}, \bibinfo{journal}{Physical Review B}
  \textbf{\bibinfo{volume}{71}}, \bibinfo{eid}{140509} (\bibinfo{year}{2005}).


\bibitem[{\citenamefont{Freeland et~al.}(2005)\citenamefont{Freeland, Gray,
  Ozyuzer, Berghuis, Badica, Kavich, Zheng, and Mitchell}}]{natmat}
\bibinfo{author}{\bibfnamefont{J.~W.} \bibnamefont{Freeland}},
  \bibinfo{author}{\bibfnamefont{K.~E.} \bibnamefont{Gray}},
  \bibinfo{author}{\bibfnamefont{L.}~\bibnamefont{Ozyuzer}},
  \bibinfo{author}{\bibfnamefont{P.}~\bibnamefont{Berghuis}},
  \bibinfo{author}{\bibfnamefont{E.}~\bibnamefont{Badica}},
  \bibinfo{author}{\bibfnamefont{J.}~\bibnamefont{Kavich}},
  \bibinfo{author}{\bibfnamefont{H.}~\bibnamefont{Zheng}}, \bibnamefont{and}
  \bibinfo{author}{\bibfnamefont{J.~F.} \bibnamefont{Mitchell}},
  \bibinfo{journal}{Nature Materials} \textbf{\bibinfo{volume}{4}},
  \bibinfo{pages}{62} (\bibinfo{year}{2005}).

\bibitem[{\citenamefont{Freeland et~al.}(2007)\citenamefont{Freeland, Kavich,
  Gray, Ozyuzer, Zheng, Mitchell, Warusawithana, Ryan, Zhai, Kodama
  et~al.}}]{jpcm}
\bibinfo{author}{\bibfnamefont{J.}~\bibnamefont{Freeland}},
  \bibinfo{author}{\bibfnamefont{J.~J.} \bibnamefont{Kavich}},
  \bibinfo{author}{\bibfnamefont{K.}~\bibnamefont{Gray}},
  \bibinfo{author}{\bibfnamefont{L.}~\bibnamefont{Ozyuzer}},
  \bibinfo{author}{\bibfnamefont{H.}~\bibnamefont{Zheng}},
  \bibinfo{author}{\bibfnamefont{J.}~\bibnamefont{Mitchell}},
  \bibinfo{author}{\bibfnamefont{M.~P.} \bibnamefont{Warusawithana}},
  \bibinfo{author}{\bibfnamefont{P.}~\bibnamefont{Ryan}},
  \bibinfo{author}{\bibfnamefont{X.}~\bibnamefont{Zhai}},
  \bibinfo{author}{\bibfnamefont{R.~H.} \bibnamefont{Kodama}},
  \bibnamefont{et~al.}, \bibinfo{journal}{J. Phys. Cond. Mat.}
  \textbf{\bibinfo{volume}{In press}} (\bibinfo{year}{2007}).

\bibitem[{\citenamefont{Chakhalian et~al.}(2006)\citenamefont{Chakhalian,
  Freeland, Srajer, Strempfer, Khaliullin, Cezar, Charlton, Dalgliesh,
  Bernhard, Cristiani et~al.}}]{natphys}
\bibinfo{author}{\bibfnamefont{J.}~\bibnamefont{Chakhalian}},
  \bibinfo{author}{\bibfnamefont{J.~W.} \bibnamefont{Freeland}},
  \bibinfo{author}{\bibfnamefont{G.}~\bibnamefont{Srajer}},
  \bibinfo{author}{\bibfnamefont{J.}~\bibnamefont{Strempfer}},
  \bibinfo{author}{\bibfnamefont{G.}~\bibnamefont{Khaliullin}},
  \bibinfo{author}{\bibfnamefont{J.~C.} \bibnamefont{Cezar}},
  \bibinfo{author}{\bibfnamefont{T.}~\bibnamefont{Charlton}},
  \bibinfo{author}{\bibfnamefont{R.}~\bibnamefont{Dalgliesh}},
  \bibinfo{author}{\bibfnamefont{C.}~\bibnamefont{Bernhard}},
  \bibinfo{author}{\bibfnamefont{G.}~\bibnamefont{Cristiani}},
  \bibnamefont{et~al.}, \bibinfo{journal}{Nature Physics}
  \textbf{\bibinfo{volume}{2}}, \bibinfo{pages}{244} (\bibinfo{year}{2006}).

\bibitem[{\citenamefont{Varela et~al.}(2003)\citenamefont{Varela, Lupini,
  Pennycook, Sefrioui, and Santamaria}}]{varela}
\bibinfo{author}{\bibfnamefont{M.}~\bibnamefont{Varela}},
  \bibinfo{author}{\bibfnamefont{A.}~\bibnamefont{Lupini}},
  \bibinfo{author}{\bibfnamefont{S.}~\bibnamefont{Pennycook}},
  \bibinfo{author}{\bibfnamefont{Z.}~\bibnamefont{Sefrioui}}, \bibnamefont{and}
  \bibinfo{author}{\bibfnamefont{J.}~\bibnamefont{Santamaria}},
  \bibinfo{journal}{Solid State Elec.} \textbf{\bibinfo{volume}{47}},
  \bibinfo{pages}{2245} (\bibinfo{year}{2003}).

\end{thebibliography}

\end{document}